# ADAPTABLE MULTI-DOMAIN LANGUAGE MODEL FOR TRANSFORMER ASR

*Taewoo Lee[1], Min-Joong Lee[2], Tae Gyoon Kang[2], Seokyeoung Jung[1], Minseok Kwon[1], Yeona Hong[1], Jungin Lee[1], Kyoung-Gu Woo[1], Ho-Gyeong Kim[2], Jiseung Jeong[2], Jihyun Lee[2], Hosik Lee[2], Young Sang Choi[2]*

[1] AI R&D Group, Samsung Electronics, South Korea
[2] Samsung Advanced Institute of Technology, Samsung Electronics, South Korea

{tw1.lee, minjoong.lee, taeg.kang, jihyun.s.lee}@samsung.com

## ABSTRACT

We propose an adapter based multi-domain Transformer based language model (LM) for Transformer ASR. The model consists of a big size common LM and small size adapters. The model can perform multi-domain adaptation with only the small size adapters and its related layers. The proposed model can reuse the full fine-tuned LM which is fine-tuned using all layers of an original model. The proposed LM can be expanded to new domains by adding about 2% of parameters for a first domain and 13% parameters for after second domain. The proposed model is also effective in reducing the model maintenance cost because it is possible to omit the costly and time-consuming common LM pre-training process. Using proposed adapter based approach, we observed that a general LM with adapter can outperform a dedicated music domain LM in terms of word error rate (WER).

*Index Terms*—End-to-end (E2E) automatic speech recognition (ASR), language model (LM), multi-domain adaptation

## 1. INTRODUCTION

In recent years, virtual voice assistants have been widely spread to real-world applications. End-to-end (E2E) automatic speech recognition (ASR) has become one of the key elements of virtual voice assistant services. As new domains continue to be added, ASR models need to be adapted quickly to the new domains. Furthermore, domain specific proper nouns must be recognized such as new song titles and singer names. This means that it is necessary to maintain the recognition accuracy of the existing supported domains while securing the recognition accuracy for new words in the new domain. In addition, in order to provide a good user experience, such a response must be done very quickly.

Transformer was first introduced as a model for translation [1]. Then, it has also been successfully applied to ASR [2]. This is because Transformer has an advantage in terms of computation and parallelism over recurrent neural network (RNN) based models. In addition, knowledge distillation has been studied to create parameter efficient models [3,4]. Shallow fusion of the E2E ASR models and external language models (LM) also showed a further improvement in WER [5,6], because external LMs are able to learn more contextual information from abundant text-only data.

In natural language processing (NLP), several methods of pre-training neural language models have led to major advances in NLP subtasks. BERT, ELMO, GPT, RoBERTa, and XLNet are typical [7-11]. These methods find dependencies between words and their combinations by pre-training neural networks on large amounts of data. Also, by fine-tuning the model on training data in target tasks, these models could be easily applied to solving other NLP tasks. However, it is difficult to continuously update these models because deep networks tend to forget previous knowledge when it is sequentially re-trained [12]. To solve such a problem, continual learning approaches have been studied. To preserve previous knowledge, learning without forgetting (LWF) [13] adds output logits of previous stage networks to logits of current stage networks. Elastic weight consolidation (EWC) [14] constrains weight updates by valuing which weight are important for a task. Progressive neural networks [15] avoid forgetting by preserving task specific networks. However, those approaches are imperfect in memory and parameter efficiency [16].

In computer vision, residual adapter modules have been introduced to make a multi-task and multi-domain model [17]. In the paper, a large common model is used as a base model. Then small adapter modules are added in front of each batch normalization layer in series or in parallel manner. In the experiments, both methods showed better accuracy than a full fine-tuned model. Similar approaches have been explored for BERT in NLP [18]. In the paper, the authors proposed a model (called projected attention layers or PALs) that can resolve multi-domain NLP tasks by adding only adjustable 13% parameters compared to the original model. Meanwhile, in [16], a method to fine-tune models by adding only adjustable 3.6% of parameters has been proposed. The method adds small size adapters to the self-attention (SA) and feed forward network (FFN) layers of Transformer, respectively. In [19], the authors compared PALs and adapters. In the paper, fine-tuning adapters with norm layer showed better results compared to the PALs when almost similar number of parameters is used. For multilingual ASR, a structure is introduced so that only adapter layers can be switched [20]. In the study, the experiments have been conducted on recurrent neural network transducer (RNN-T) based streaming E2E ASR models.

In this paper, we study an external LM structure for Transformer based ASR model that can be adapted for multi-domain with only 2% or 13% parameter addition per domain. To the best of our knowledge, this is a first attempt applying adapters to Transformer LM in ASR. The effects of our model are: 1) Our adapter-based adaptation can be used on top of the full fine-tuned model, and it further reduces word error rate (WER) from the model. 2) Multi-domain LM can be supported with fewer parameters. 3)

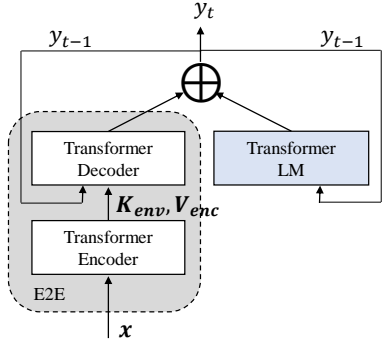

**Fig. 1.** The dotted line box shows transformer-based E2E ASR model, including encoder and decoder. An external LM is incorporated at each step of beam search.

Our approach provides cost efficient way to maintain existing models.

## 2. SA-BASED MULTI –DOMAIN LM WITH ADAPTER

### 2.1. Transformer-based E2E ASR

Figure 1 shows a Transformer based E2E ASR models with an external LM. As in [2], the encoder module, which is similar to an acoustic model, takes the input features, $x$, and transforms them to a higher-level feature representation with self-attention layers. The outputs of the encoder key $K_{enc}$ and value $V_{enc}$ are passed to encoder-decoder attention layers of E2E decoder. Using the $K_{enc}$ and $V_{enc}$, the E2E decoder iteratively predicts output probabilities $P(y_t|y_0,\cdots,y_{t-1},x)$ of next output symbol $y_t$ until maximum sequence length or EOS (end-of-sequence) is met. An external LM [25], where encoder-decoder attention layers are removed, can be incorporated at each step of beam search to improve accuracy. Hereafter, we focus on an external LM decoder with adapters.

### 2.2. SA-based LM Decoder with Adapter

SA-based LM decoder consists of three parts: an input embedding, $N_L$ LM SA layers, and a linear transform following Softmax (Figure 2 left). For simplicity we set batch size and the number of domains to one in the followings.

#### 2.2.1 Input Embedding

Let word-piece [21] vocabulary size be $N_w$, an input one-hot vector be $x_t \in \mathbb{R}^{1 \times N_w}$, hidden size be $h$. The output of embedding matrix is computed as (1):

$$W_e^o = x_t W_e \quad (1)$$

where $W_e \in \mathbb{R}^{N_w \times h}$ and $W_e^o \in \mathbb{R}^{1 \times h}$. Then a positional encoding vector $PE \in \mathbb{R}^{1 \times h}$ is added to $W_e^o$ [1].

#### 2.2.2. SA layer in LM Decoder with Adapter

A SA layer of a LM decoder with adapters consists of four layers: layer norm [22], multi-head attention (MHA), FFN, and adapters.

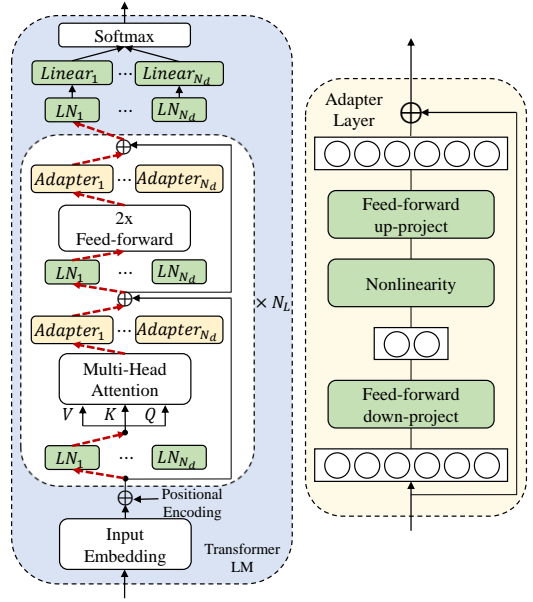

**Fig. 2.** (Left) is an architecture of transformer multi-domain LM. In a LM decoder, the adapter module (right) is added on top of multi-head attention and feed-forward layers [16]. Only green layers (including layer norms or LN) are fine-tuned on the downstream data and expanded for $N_d$ multi-domain. Dotted red lines shows a switchable decoding path for a first domain.

##### 2.2.2.1 Multi-Head Attention

Let the number of heads be $N_{head}$. Previous output is projected to a query, a key, and a value simultaneously for multi-head attention (Figure 2 left). Instead of performing a single attention function using $h$ dimentional $Q$, $K$, and $V$, MHA performs the attention function $N_{head}$ times in parallel with differently learned $h/N_{head}$ dimentional $Q$, $K$, and $V$. Then $N_{head}$ numbers outputs are concatenated and projected into a single representation. The detailed equation is as follows:

$$\text{MultiHead}(Q,K,V) = \text{Concat}(head_1,\cdots,head_{N_{head}})W^O \quad (2)$$

$$\text{where} \quad head_i = Attention(QW_i^Q, KW_i^K, VW_i^V) \quad (3)$$

$$= \text{softmax}\left(\frac{(QW_i^Q)(KW_i^K)^T}{\sqrt{\frac{h}{N_{head}}}}\right)(VW_i^V),$$

$W_i^Q \in \mathbb{R}^{h \times d_q}$, $W_i^K \in \mathbb{R}^{h \times d_k}$, $W_i^V \in \mathbb{R}^{h \times d_v}$, and $W^O \in \mathbb{R}^{h \times h}$ are trainable parameters. Note $d_q = d_k = d_v = h/N_{head}$ throughout the paper.

##### 2.2.2.2 Position-wise Feed-Forward Network

Let an inner filter size $f$. Position-wise feed forward network consists of two FFNs with ReLU activation in between. An output of position-wise FFN is calculated as (4) where the input vector $i_1 \in \mathbb{R}^{1 \times h}$, the weight matrices and bias vectors $W_1 \in \mathbb{R}^{h \times f}$, $b_1 \in \mathbb{R}^{1 \times f}$, $W_2 \in \mathbb{R}^{f \times h}$, and $b_2 \in \mathbb{R}^{1 \times h}$.

**Table 1.** The architectures and sizes of SA E2E, general LM (G-LM), music LM (M-LM), and adapter added LMs

|  | E2E Enc. | E2E Dec. | G-LM | G-LM-A | M-LM | M-LM-A |
|---|---|---|---|---|---|---|
| # Layers | 6 | 4 | 3 | 3 | 2 | 2 |
| $h$ | 512 | 512 | 512 | 512 | 512 | 512 |
| $f$ | 3072 | 3072 | 4096 | 4096 | 2048 | 2048 |
| $f_A$ | - | - | - | 64 | - | 64 |
| $N_{head}$ | 16 | 4 | 8 | 8 | 8 | 8 |
| Size (MiB) | 96.7 | 80.3 | 76.4 | 77.9 | 40.3 | 41.3 |

$$\text{FFN}(i_1) = \max(0, i_1 \boldsymbol{W}_1 + b_1) \boldsymbol{W}_2 + b_2 \quad (4)$$

*2.2.2.3 Adapter*

Adapter modules proposed in [16] are inserted on top of MHA and FFN layers as in Figure 2 (left). An adapter module (Figure 2 right) consists of two linear transforms and ReLU activation in between. A residual connection is added to the output. The outputs of adapters $A_1$ and $A_2$ are calculated as follows:

$$A_1(i_2) = i_2 + \max(0, i_2 \boldsymbol{W}_3 + b_3) \boldsymbol{W}_4 + b_4 \quad (5)$$
$$A_2(i_3) = i_3 + \max(0, i_3 \boldsymbol{W}_5 + b_5) \boldsymbol{W}_6 + b_6, \quad (6)$$

where $i_2 = \text{MultiHead}(Q, K, V)$, $i_3 = \text{FFN}(i_1)$, adapter filter size is $f_A$, $\boldsymbol{W}_3, \boldsymbol{W}_5 \in \mathbb{R}^{h \times f_A}$, $b_3, b_5 \in \mathbb{R}^{1 \times f_A}$, $\boldsymbol{W}_4, \boldsymbol{W}_6 \in \mathbb{R}^{f_A \times h}$, $b_4, b_6 \in \mathbb{R}^{1 \times h}$.

*2.2.3 Softmax*

The outputs of decoder are transformed to the probabilities of output classes by a linear projection $\boldsymbol{W}_7 \in \mathbb{R}^{h \times N_w}$ and a subsequent softmax function.

## 3. EXPERIMENTS

Table 1 shows overall model architectures and model sizes used in the experiments. In the experiments, a general domain LM (G-LM), a music specialized domain LM (M-LM), and adapter added general and music LMs (G-LM-A, M-LM-A) are used. For single precision floating point, model sizes are increased about 2% when adapters are added for a first domain.

The G-LM is trained on 24GiB normalized Korean text data consisting of 353M utterances. All data were anonymized. The data consists of representative utterances of Samsung's Bixby scenario and general domain corpus. The M-LM is trained on normalized Korean text data consisting of 45M utterances, in which general and music domain (song title and singer name related commands) corpus are mixed. To train our models, we used Tensor2Tensor framework [23].

For G-LM experiments, we recorded test cases (TCs) in three categories: In-Domain, Out-Domain, and Open-Domain. In-domain TCs are having a similar data distribution with training cases but out-domain TCs are not. To ensure the difference, Out-Domain TCs are not only from completely different domain (e.g. doctor-patient conversation) but also include unique proper nouns. Open domain TCs have been added to confirm that there is no performance degradation before and after adaptation. In-Domain TCs includes 50K Bixby use-case scenario utterances such as phone and device control commands and daily conversational question and answering. Out-Domain TCs includes 8K domain specific utterances which is not included in In-Domain training corpus. Especially, we selected domains having its own unique proper nouns such as hospital or doctor's names. Open-Domain TCs are included to test noisy environment, on which cafe, city, office, highway noises are added to clean speech. The content of the utterances is in arbitrary domain and do not include unknown unique proper nouns. All TCs are recorded in male and female voices. For M-LM experiments, In-Domain and Out-Domain TCs are recorded. In-Domain TC includes 610 utterances. It represents well known song titles and singer names. On the other hand, Out-Domain TC includes 3709 utterances. The content is newly added song titles and singer names.

**Table 2.** WERs of E2E, E2E-G-LM, and E2E-G-LM-A on General Domain TCs

| TC | E2E | E2E-G-LM | E2E-G-LM-A |
|---|---|---|---|
| In-Domain | 2.42 | 1.82 | 1.69 |
| Out-Domain | 10.62 | 8.18 | **2.84** |
| Open-Domain | 12.8 | 5.08 | 4.55 |

**Table 3.** WERs of E2E, E2E-M-LM, and E2E-M-LM-A on Music Domain TCs

| TC | E2E | E2E-M-LM | E2E-M-LM-A |
|---|---|---|---|
| In-Domain | 8.2 | 2.68 | 2.46 |
| Out-Domain | 12.66 | 5.43 | **4.13** |

**Table 4.** WERs of iterative adapter fine-tuning with M-LM-A on Music Domain TCs

| TC | E2E-M-LM | M1$_{iter1}$ | M1$_{iter2}$ | M1$_{iter3}$ |
|---|---|---|---|---|
| In-Domain | 2.68 | 2.46 | 1.97 | **1.81** |
| Out-Domain | 5.43 | 4.13 | 3.96 | **3.87** |

We initialized weights of each adapter layer to the values following a normal distribution having zero mean and $10e^{-4}$ variance. We tested variance values of $\{0, 10e^{-7}, 10e^{-6}, 10e^{-5}, 10e^{-4}, 10e^{-3}, 10e^{-2}\}$ and selected a largest stable value. Since an adapter module internally has a residual connection, zero variance can be inserted to test output of the adapter module is bypassed properly. All runs are trained on eight P40 GPUs to build models from scratch and on one P40 GPU for adaptations. We used Adam optimizer with $\beta_1 = 0.9$, $\beta_2 = 0.98$, $\epsilon = 1e^{-9}$. Batch sizes tested from $\{32, 64, 128, 512, 1024, 4096, 8192\}$. 8192 is used for all our adaptation experiments. Unlike [16], small batch size made our training unstable, failing to converge. Learning rate is selected as 0.03 from $\{0.1, 0.03, 0.001, 0.0003, 0.0001\}$. When we train our models from scratch or adapt without adapter, we applied Noam learning rate decay scheme with 1000 warmup steps. On the other hand, when we train our adapter related layers, learning rate decay scheme did not used.

**Table 5.** Iterative fine-tuning performance (WER). The results show a G-LM with iterative fine-tuned adapters can be used as a dedicated music LM.

| TC | E2E-M-LM | E2E-M-LM-A | E2E-G-LM | E2E-G-LM-$A_{iter1}$ | E2E-G-LM-$A_{iter2}$ | E2E-G-LM-$A_{iter3}$ | WERR (E2E-G-LM-$A_{iter3}$ - E2E-M-LM) | WERR (E2E-G-LM-$A_{iter3}$ - E2E-M-LM-A) |
|---|---|---|---|---|---|---|---|---|
| In-Domain | 2.68 | 2.46 | 4.65 | 3.82 | 2.38 | 2.19 | -0.49 | -0.27 |
| Out-Domain | 5.43 | 4.13 | 11.27 | 5.75 | 4.75 | 4.60 | -0.83 | 0.47 |

We used 4096 word-pieces as output token units. For E2E model training, we used same hyper-parameters in [3]. All experiments used the identical input feature processing to that of [24]. The decoding hyper-parameters (beam size = 4, length-penalty = 1.2, and maximum decoding length = 80) were tuned to minimize WER. Known proper nouns and number are converted with an inverse text normalization (ITN) module. We assumed we already knew proper domain names before inferencing.

## 4. RESULTS

Table 2 shows WERs measured with only E2E models (E2E), E2E models with a full fine-tuned G-LM (E2E-G-LM), and E2E models with an adapter fine-tuned G-LM (E2E-G- LM-A). Compared to the results decoded with only E2E models (E2E), in E2E-G-LM, WERs were reduced 0.6, 2.44, and 7.72%p for in, out, and open domain TCs, respectively. When the full fine-tuned G-LM was additionally adapter fine-tuned (E2E-G-LM-A), WERs were further reduced by 0.73, 7.78, and 8.25%p for in, out, and open domain TCs respectively. In particular, in domains having unusual proper nouns, we got higher improvement in accuracies. This means adapter fine-tuning can bias output probability properly for unusual proper nouns. In addition, despite this strong biasing, the accuracy of existing domain TCs did not deteriorated.

Table 3 shows WERs measured with only E2E models (E2E), E2E models with a full fine-tuned M-LM (E2E-M-LM), and E2E models with an adapter fine-tuned M-LM (E2E-M- LM-A). The results of using E2E models with a full fine-tuned M-LM (E2E-M-LM) showed improved WERs than the results decoded with the E2E models alone. The WERs of in and out domain TCs were reduced by 5.52 and 7.23%p, respectively. When the full fine-tuned M-LM was additionally adapter fine-tuned (E2E-M-LM-A), WERs were further reduced by 0.22, 1.3%p for in and out domain TCs respectively. Like G-LM experiments, adapter fine-tunings improves the proper noun recognition accuracy without compromising the accuracy of existing domains, even for smaller models.

In Table 4, we see how far WERs can be reduced by iterative adapter fine-tuning. The model M1 refers to a model that an adapter fine-tuned M-LM using error sentences from the E2E-M-LM decoding result as training data. We considered decoding, error sentence extraction, and re-training a model as one iteration. From a model training point of view, an iteration also can be defined to find a point where loss values are minimal when we train with a small amount of iterative adaptation data. In the experiment, accuracy improved until iterations were repeated three times. Although the loss value was a minimum in 4th iterative adaptation, the WER increase was observed. Therefore, the training has been stopped at the point based on WER.

Table 5 compares the case of using a G-LM as a common base LM with an iterative adapter fine-tuned G-LM (E2E-G-$LM_{iter}$) and the case of creating a dedicated M-LM and full fine-tune or adapter fine-tune it (E2E-M-LM, E2E-M-LM-A). Intuitively, when we decode music domain TCs with the E2E model and G-LM without any adaptation (E2E-G-LM) as a baseline, it showed a higher error rates than E2E-M-LM and E2E-M-LM-A. Last two columns in Table 5 show word error rate reduction (WERR). When we iterative adapter fine-tuned the G-LM three times (E2E-G-LM-$A_{iter3}$), WERs were reduced by 0.49 and 0.83%p in both in and out domain TCs, respectively, compared to E2E-M-LM. Also, the WERs of E2E-G-LM-$A_{iter3}$ were almost close to the results of E2E-M-LM-A. This means that a common G-LM with adapters can be used as a dedicated domain LM, and we can switch only adapter related layers to fit our model on each domain. Therefore, a multi-domain LM configuration with the structure shown in Figure 2 is possible.

Since $f_A$ is a relatively small value, the increasing number of parameters per domain is about 2% for the first domain and about 13% for after the second domain. Specifically, $2N_L(2f_A h + f_A + h)$ for the first domain, because norms and Softmax linear layers can be reused. $2N_L(2f_A h + f_A + 3h) + (2h + hN_w + N_w)$ for after the second domain. This slow increasing property is important because memory size is limited for GPU or on-device applications.

We built our base LMs from scratch on eight P40 GPUs and on v3-8 tensor processing units (TPU). It took three days on eight P40 GPUs and 4 hours and 30 minutes on TPU. Iterative adapter fine-tuning proposed in the paper can train G-LM in 60 minutes on a P40 GPU and 25 minutes for M-LM. Since a P40 GPUs may be available in on premise servers, we expect that cloud computing cost will be saved.

## 5. CONCLUSIONS

In this paper, adapter based multi-domain LM structure has been proposed. The structure is a combination of two architectures: an adapter module proposed for BERT in NLP area and a switchable adapter architecture proposed for RNN-T streaming ASR model. The proposed architecture allows LMs to expand multi-domain, suppressing the increase of the number of parameters. The proposed architecture can reduce WERs of target domains without WER decrease of existing domains. Also we observed that applying adapter module on Transformer LM has an effect on WER improvement especially for proper nouns that is hard to be handled with a common base LM. Finally, the proposed architecture can reuse standard full fine-tuned LMs. So, the full fine-tuned LMs can be easily reused (or transferred) without any changes.